\documentclass[journal, twocolumn]{IEEEtran}

\usepackage{graphicx}
\graphicspath{{./figures/}}
\DeclareGraphicsExtensions{.pdf}
\usepackage{epstopdf}
\usepackage{subcaption}
\usepackage{multirow}
\usepackage{amsmath}
\usepackage[numbers]{natbib}
\usepackage{url}

\begin{document}

\title{Velocity determination for time lapse surveys using residual prestack depth migration}

\author{Joerg F. Schneider \\
  Bureau of applied Geophysics, Nordstemmen, Germany}

\maketitle

\begin{abstract}
  The presented approach aims at estimating the lateral variation of seismic velocities for a seismic timelapse survey. The monitor survey is depth migrated with the known velocity model of the base survey. An analysis is performed to determine whether the moveout to be expected for the monitor survey after residual prestack depth migration (RPSM) can be used for the velocity determination of the formation of interest. In that case the migrated response for the base horizon of the monitor survey is analyzed in order to determine the velocity distribution by employing residual migration concepts. With the use of Fresnel apertures a rather local application is possible at comparatively low signal to noise ratios.
The approach is successfully demonstrated for two isotropic depth models both for a lenslike and for a faulted reservoir structures. In both cases the velocities of the monitor survey is recovered accurately. In addition it is demonstrated that an iterative application is possible and that for these two models even severe compaction of the model has little influence on the velocity estimation.   
\end{abstract}

\section{Introduction}
Prestack migrations (PSM) have been performed for decades for the processing of seismic surveys either as prestack time (PSTM) or prestack depth (PSDM) migrations (Yilmaz, 2001, Bednar, 2005, Robein 2010). Previously, but also in recent years (Douma and de Hoop, 2006) the migration and demigration of interpreted reflection responses has been considered (Lambare et alii, 2008, Guillaume et alii, 2008).\\The presented approach aims at estimating the lateral variation of seismic velocities for a seismic timelapse survey. The monitor survey is depth migrated with the known velocity model of the base survey. An analysis is performed to determine whether the moveout to be expected for the monitor survey after residual prestack depth migration (RPSM) can be used for the velocity determination of the formation of interest. In that case the migrated response for the base horizon of the monitor survey is analyzed in order to determine the velocity distribution by employing residual migration concepts.

In (Schneider, 2020) a different approach was suggested: after PSM migrated reflection responses are estimated from residual moveout (RMO) analyses. The migrated responses are demigrated and migrated into a new velocity model by applying aplanatic constructions: according to (Hagedoorn, 1954) the migrated response can be obtained as the envelope of all surfaces of equal reflection times of the source receiver pairs. This approach was used in (Schneider, 2021) to consider expanding waves along zero offset rays in order to predict residual migrations for up to intermediate offsets. Here it will be attempted to illustrate an application which is not restricted in offsets.

\section*{Method}

Both surveys of a timelapse survey are migrated with the velocity model which is assumed to be correct for the base survey. A main problem in the velocity determination for the monitor survey is the decrease in residual moveout with increasing depth and decreasing width of the formation (Schneider, 2014).\\
It is difficult to predict this quantity and its vertical and lateral resolution for a general depth model. For a prediction and in order to assess the applicability and sensitivity with respect to relevant inversion parameters recently suggested RPSM techniques are employed (Schneider, 2021): lateral and vertical variations of the isotropic or anisotropic velocity in the reservoir/region of interest are possible. For an anomaly it is assumed that at least the velocity behavior outside the anomalous region is known. For the migrated base survey and the expected formation velocity for the monitor survey an inverse RPSM is performed where for the final migration the velocities of the base survey are used. The resulting section is an approximation to the migrated response of the monitor survey; the computations are performed both for a synthetic section and possibly also for the actual base survey, in order to test the approach under different signal to noise conditions. With these two synthetic sections the algorithm is applied: again Fresnel aperture (Born, 1985, Schneider, 1989, Lindsey, 1989) RPSM and horizon analyses techniques are used in connection with semblance calculations (Taner and Koehler, 1969, Khoshnavaz, Siahkoohi and Kahoo, 2021) and optimization approaches such as Powell’s BOBYQA method (Powell, 2009) to estimate the lateral changes. The influence of compression in the overburden can be estimated by changing the velocity model of the monitor survey appropriately (as will be demonstrated). If meaningful results are obtained, one can proceed with the application to the migrated monitor survey.\\
The presented approach has the advantage that with the use of Fresnel apertures a rather local application is possible at comparatively low signal to noise ratios.

\section*{Applications}

\begin{figure}
\captionsetup{justification=justified, singlelinecheck=off}
\centering
\includegraphics[width=0.45\textwidth]{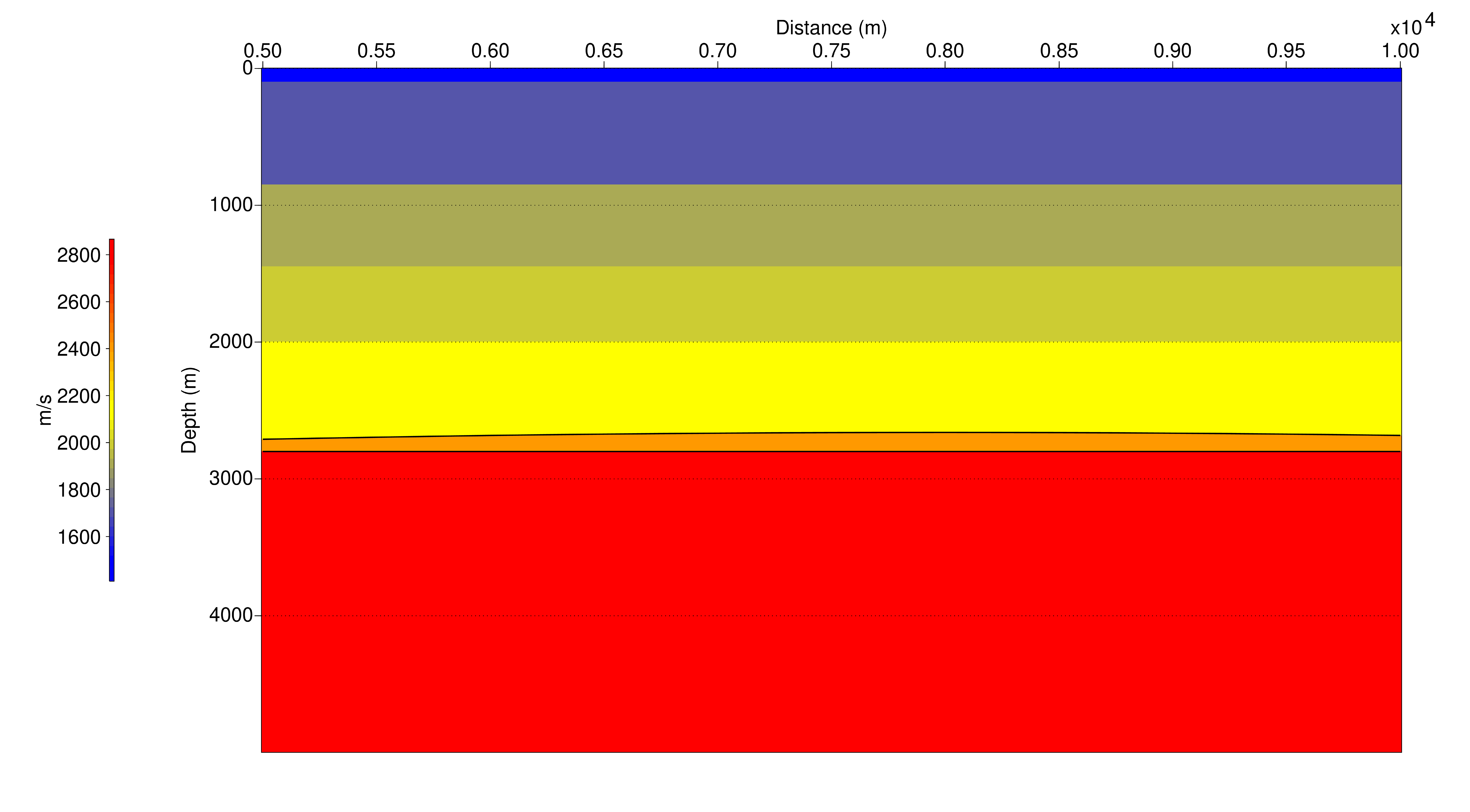}
\caption{velocity model for a reservoir structure (base model), with top and base indicated in black }
\label{fig:fig1}
\end{figure}

Figure~\ref{fig:fig1} shows the depth model and the P-wave velocities for the base survey featuring some resemblance to the lens shaped reservoir model described in (Roste et alii, 2006), here with a maximum reservoir thickness of approximately 150m.

\begin{figure}
\captionsetup{justification=justified, singlelinecheck=off}
\centering
\includegraphics[width=0.45\textwidth]{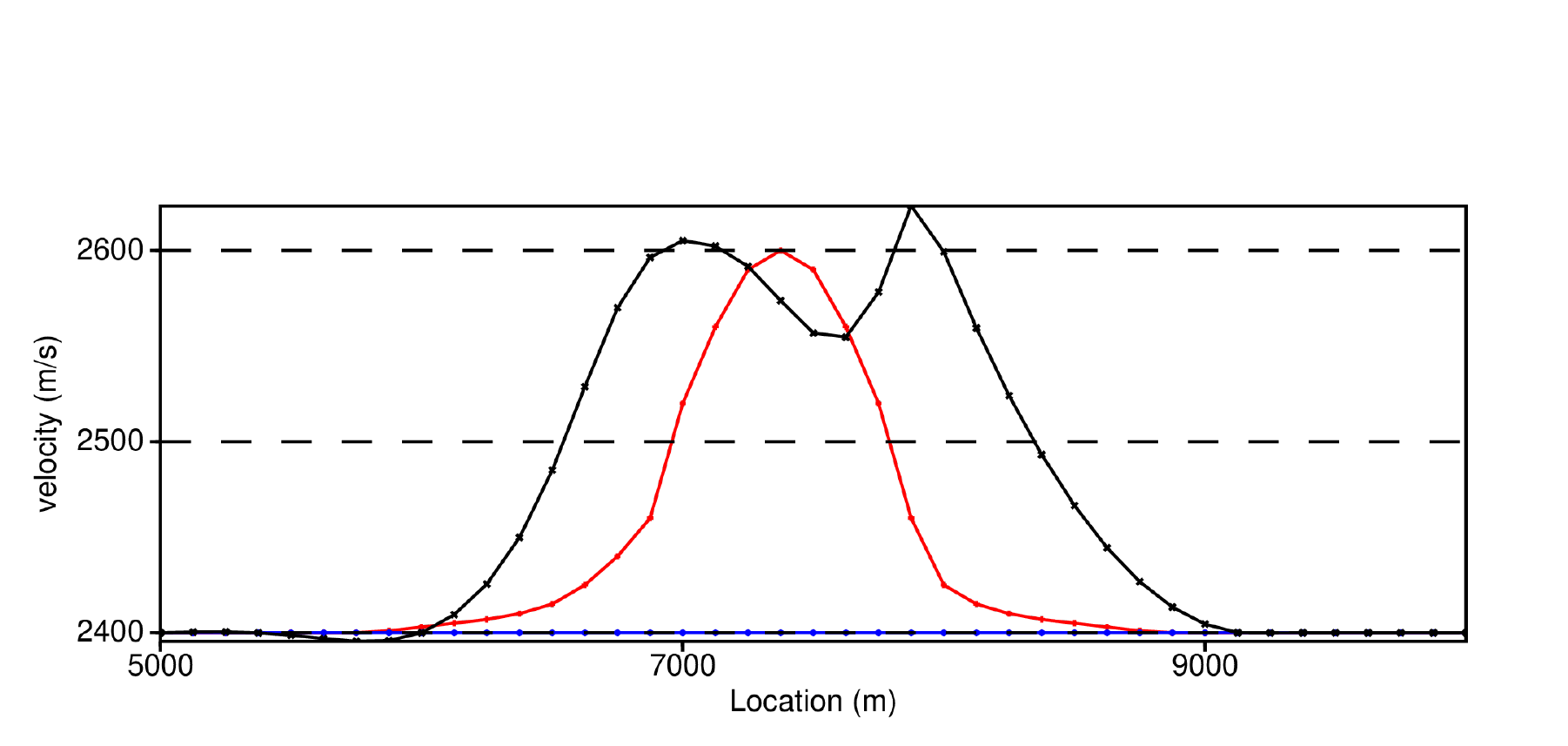}
\caption{reservoir velocities(monitor survey), true (red), initial (blue) and computed (black) }
\label{fig:fig2}
\end{figure}

\begin{figure}
\captionsetup{justification=justified, singlelinecheck=off}
\centering
\includegraphics[width=0.45\textwidth]{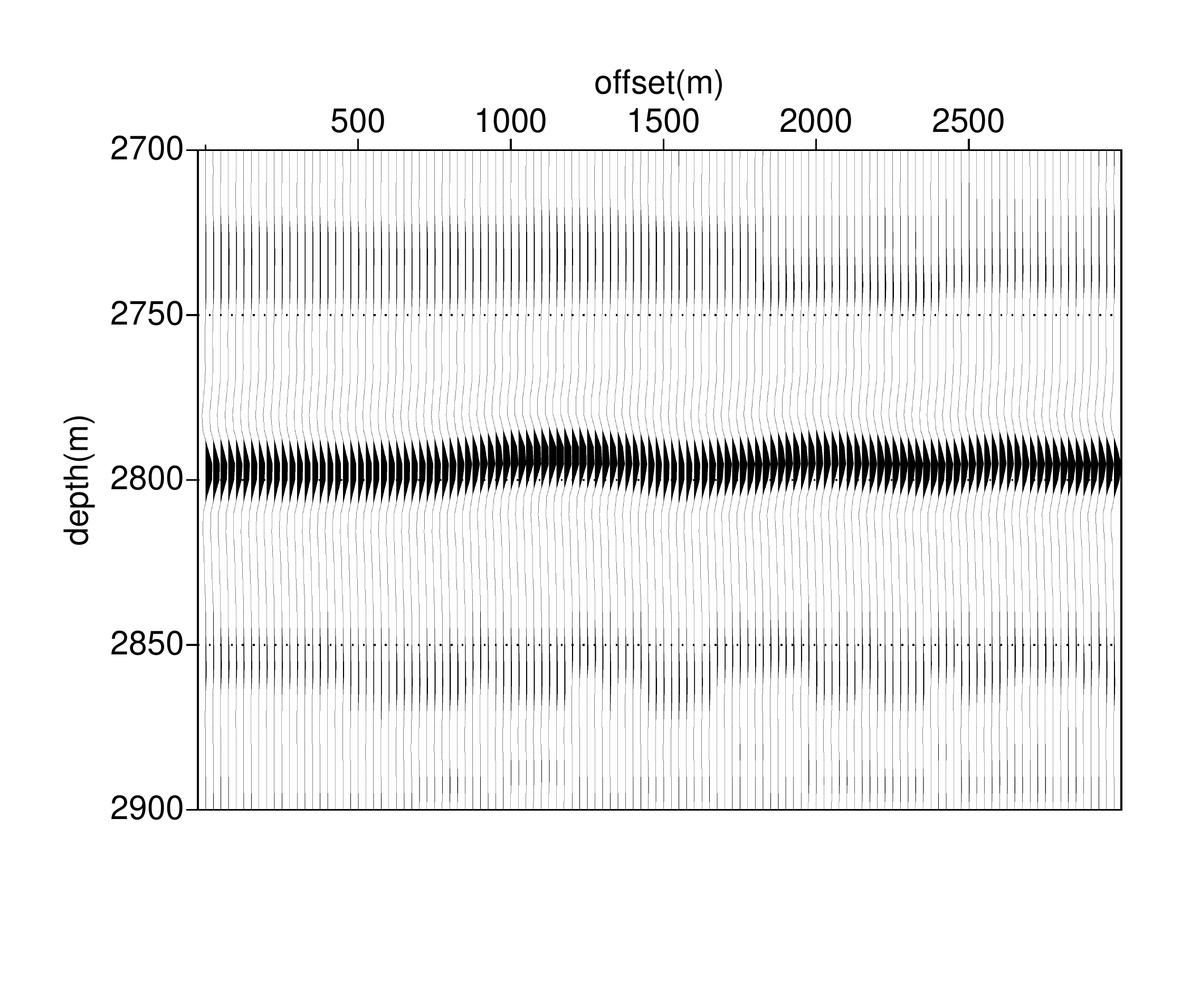}
\caption{CDP gather (monitor survey) at 8000m, migration with base model velocities}
\label{fig:fig3}
\end{figure}

\begin{figure}
\captionsetup{justification=justified, singlelinecheck=off}
\centering
\includegraphics[width=0.45\textwidth]{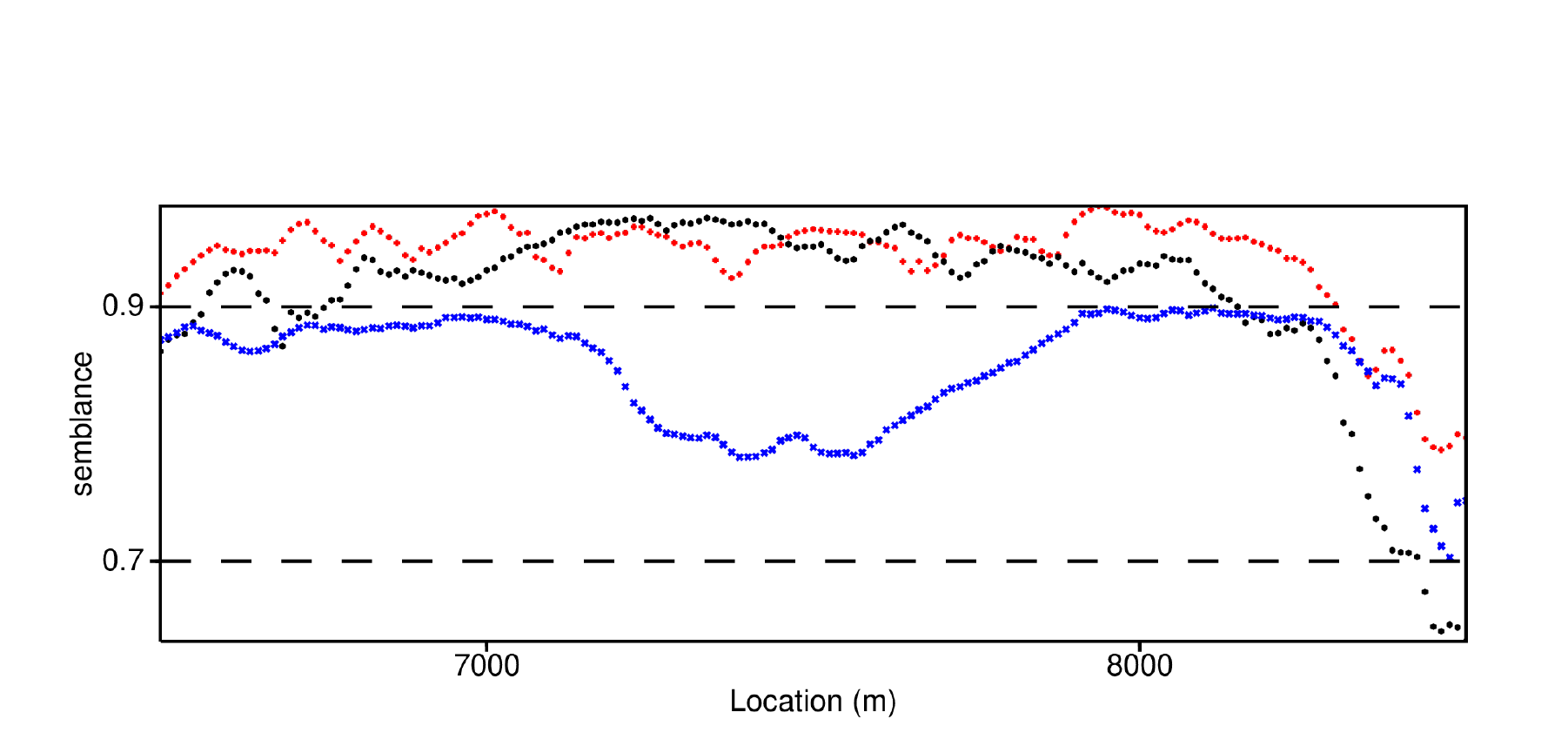}
\caption{semblances for reservoir base, with true (red), initial (blue) and computed (black) velocities}
\label{fig:fig4}
\end{figure}

\begin{figure}
\captionsetup{justification=justified, singlelinecheck=off}
\centering
\includegraphics[width=0.45\textwidth]{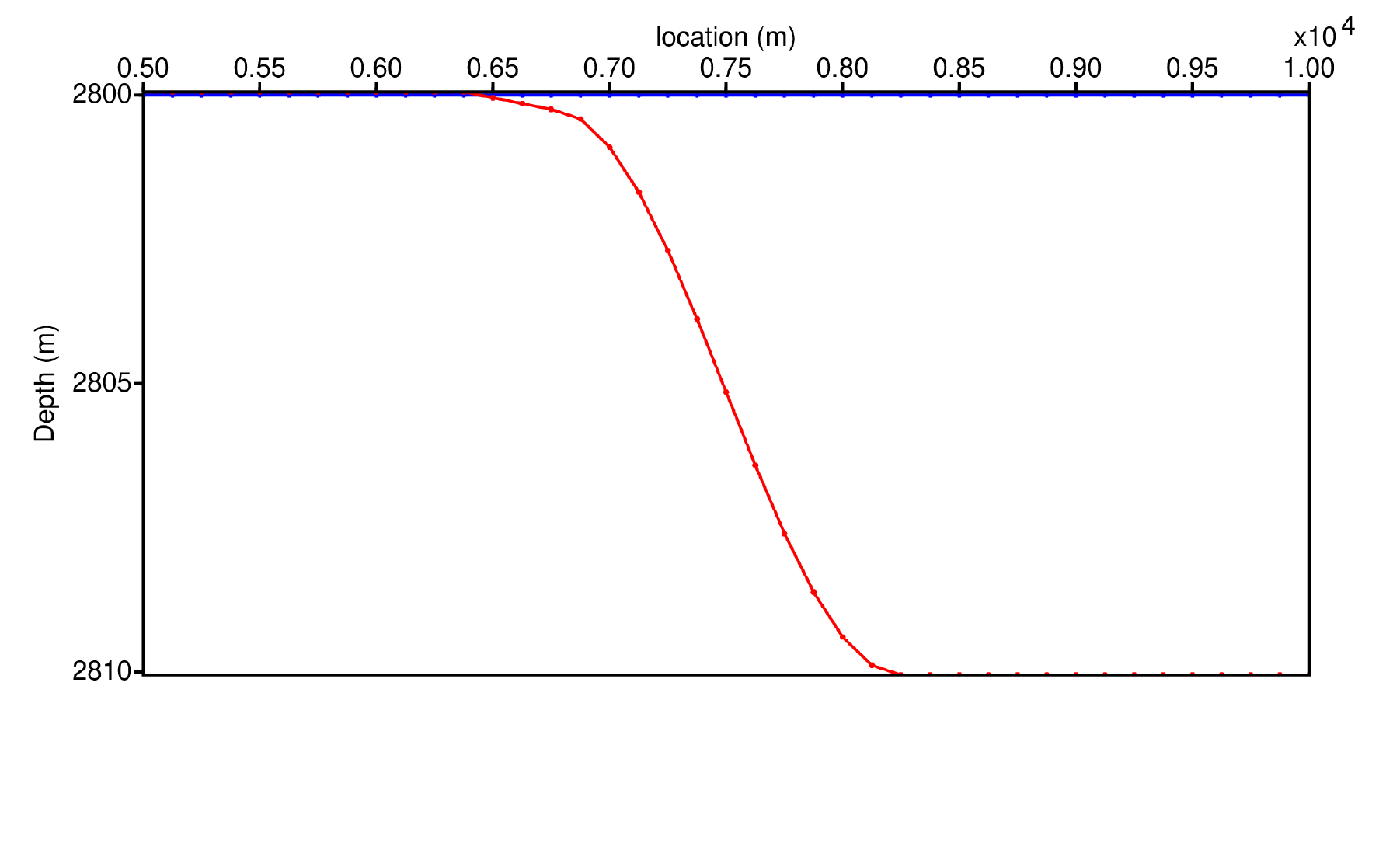}
\caption{compaction for the reservoir base: initial position (blue) and after compaction (red)}
\label{fig:fig5}
\end{figure}

\begin{figure}
\captionsetup{justification=justified, singlelinecheck=off}
\centering
\includegraphics[width=0.45\textwidth]{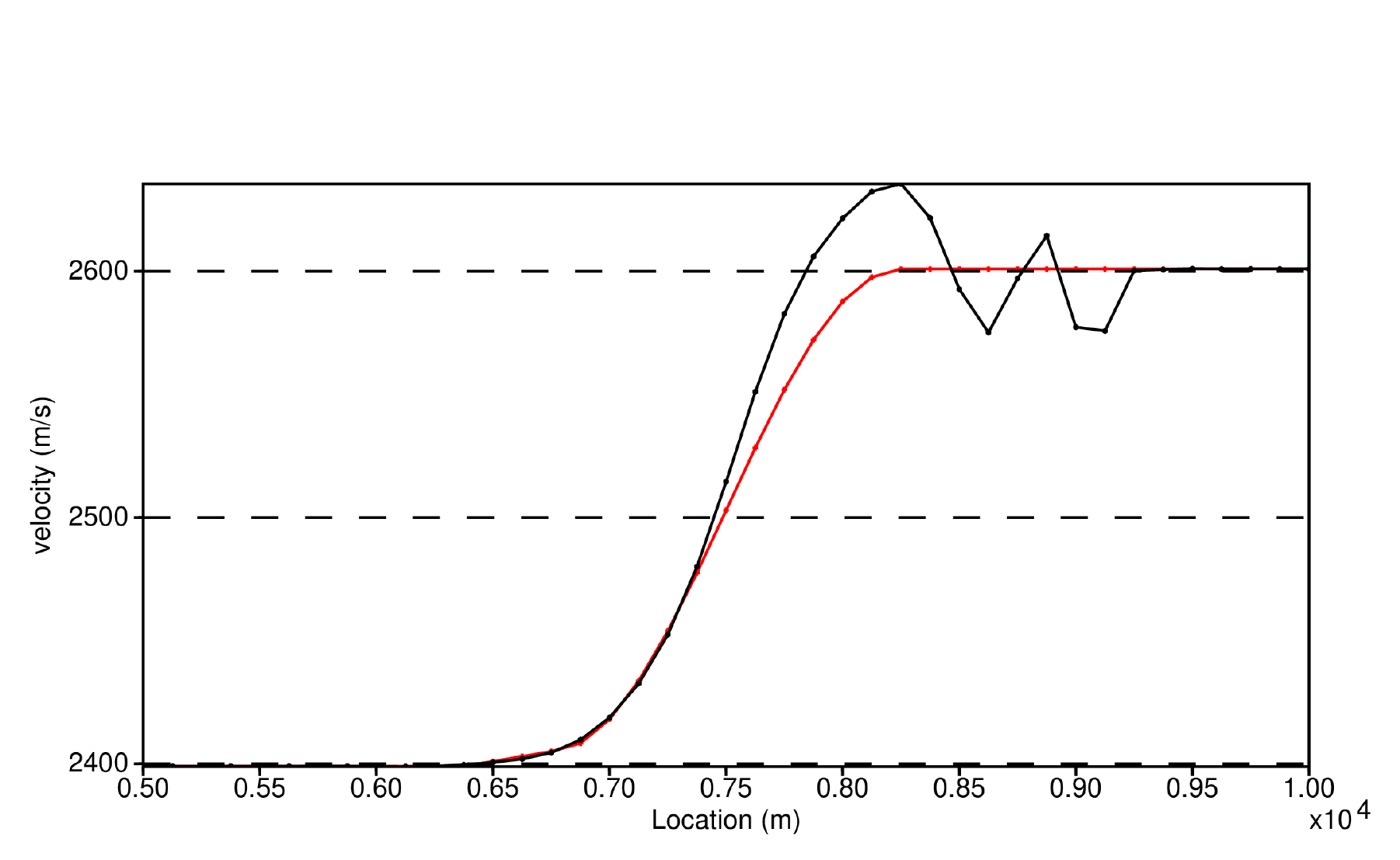}
\caption{reservoir velocities (monitor survey), true (red), computed (black) and computed after compaction (black)}
\label{fig:fig6}
\end{figure}

Two seismic surveys were assumed to be performed with 200 shots, 120 receiver/shot, maximum offset: 3000m.  The variation of the reservoir velocity for the monitor survey is shown in Figure~\ref{fig:fig2} in red, whereas the survey was migrated with the velocity model of the base survey (for illustrative purposes the reservoir velocity was assumed to be constant here). Figure~\ref{fig:fig3} shows the traces of this migration at location 8000m calculated by employing constant offset summation techniques; it can be observed that for this location the maximum residual moveout occurs at intermediate offsets. The constant velocity of the base model (2400 m/s) is presented by the blue curve in Figure~\ref{fig:fig2}; it was used as starting model fo the velocity inversion. The semblances of this erroneous migration as compared to the results obtained with the correct velocities and to the computed reservoir velocities are presented in Figure~\ref{fig:fig4} in blue, red and black, respectively. As stated above it was assumed that for the monitor survey the velocity variation was known away from the anomalous region. The velocity determination was applied as described with the difference that the approach was applied for a few iterations, where the computed velocities were used as initial distribution for the subsequent iteration. The final distribution is presented in Figure~\ref{fig:fig2} (black) and is in fair agreement with the true velocity.\\
In Figure~\ref{fig:fig5},~\ref{fig:fig6} the sensivity of the computation with respect to reservoir compaction is demonstrated for the same model with a different velocity configuration (Figure~\ref{fig:fig6}). Figure~\ref{fig:fig5} shows an artificial displacement of the reservoir base proportional to the velocity variation of the monitor survey with a maximum value of 10m. Four horizon boundaries in the overburden were displaced with similar distributions with maximum displacement values of 8m, 6m, 4m and 2m, respectively (starting from the reservoir top).  These displacements were assumed to be unknown, the uncompressed velocity model was used for the velocity estimation. The computed result for the uncompressed model in Figure~\ref{fig:fig6} (black) shows a lateral deviation with respect to the true distribution in the vicinity of the upper velocity plateau (outside the computational range $7000m < x < 8500m$ the velocities were extended  by values of 2400m/s and 2600m/s, respectively). A seperate model calculation was performed for the compression; the computed velocities in Figure~\ref{fig:fig6} were found to be identical for both velocity models.\\

\begin{figure}
\captionsetup{justification=justified, singlelinecheck=off}
\centering
\includegraphics[width=0.45\textwidth]{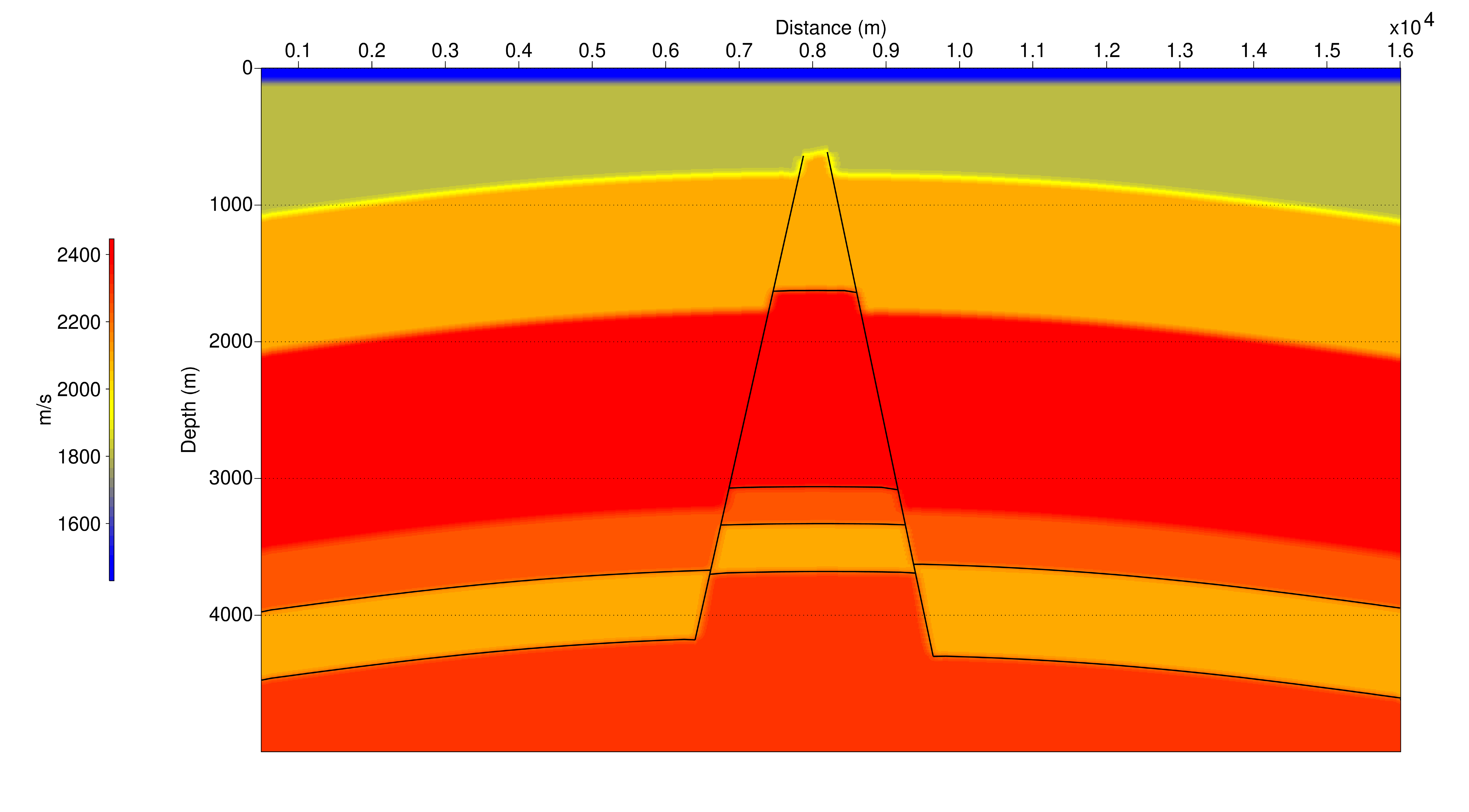}
\caption{velocity model for a reservoir structure (base model, some boundaries in black)}
\label{fig:fig7}
\end{figure}

\begin{figure}
\captionsetup{justification=justified, singlelinecheck=off}
\centering
\includegraphics[width=0.45\textwidth]{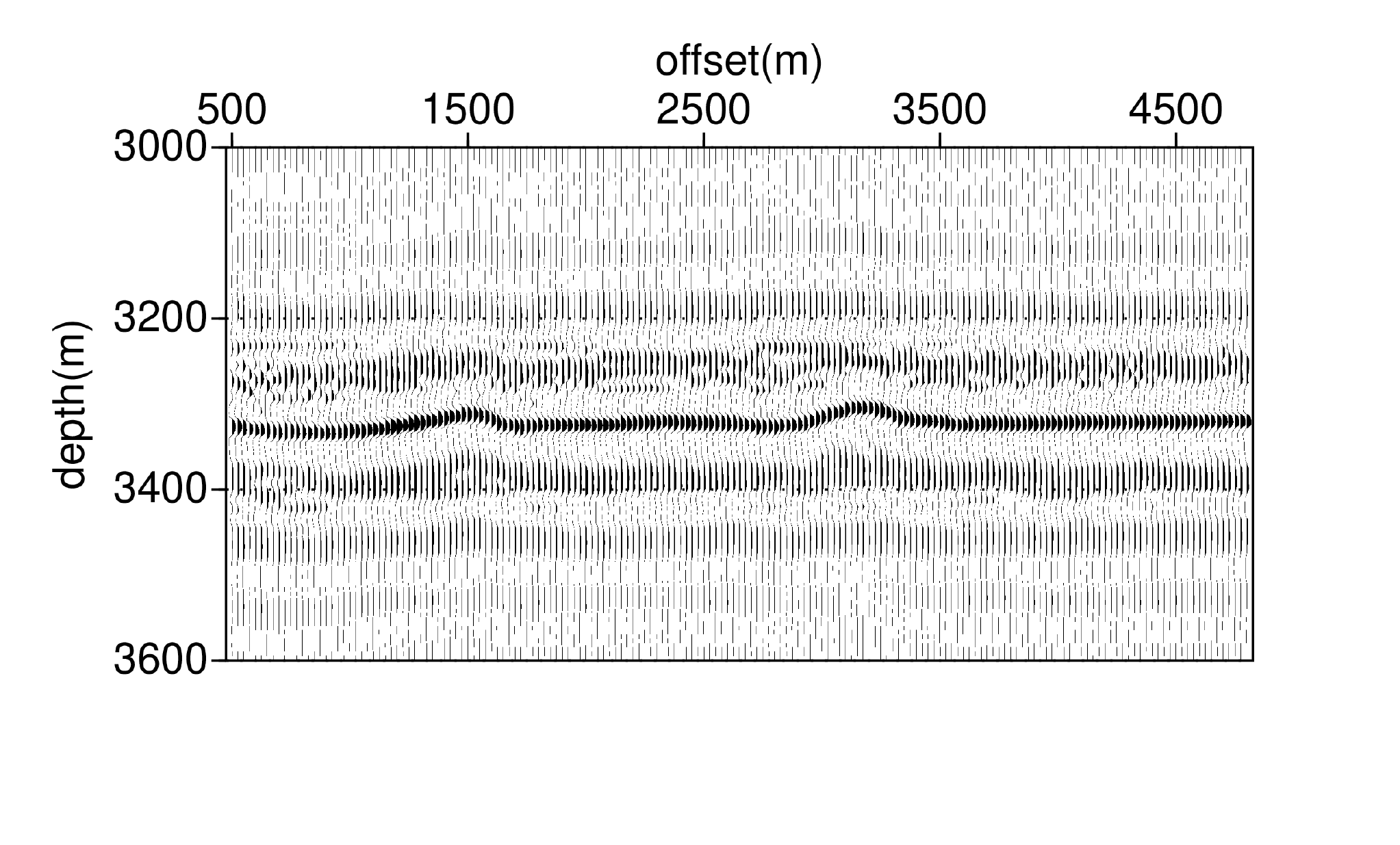}
\caption{CDP gather (monitor survey) at 7800m after migration with initial velocities }
\label{fig:fig8}
\end{figure}

\begin{figure}
\captionsetup{justification=justified, singlelinecheck=off}
\centering
\includegraphics[width=0.45\textwidth]{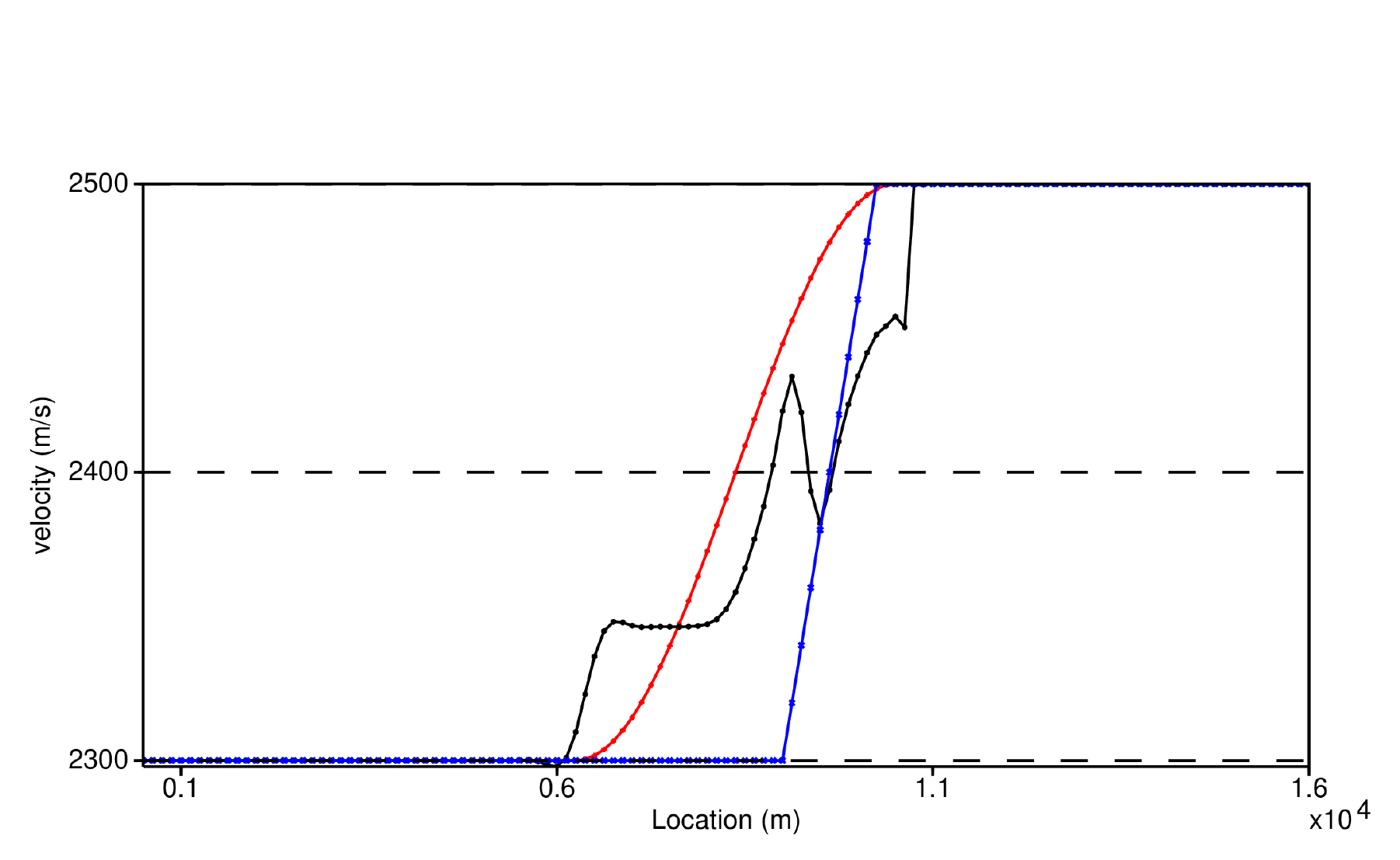}
\caption{reservoir velocities (monitor survey), true (red), initial (blue), computed (black)}
\label{fig:fig9}
\end{figure}

\begin{figure}
\captionsetup{justification=justified, singlelinecheck=off}
\centering
\includegraphics[width=0.45\textwidth]{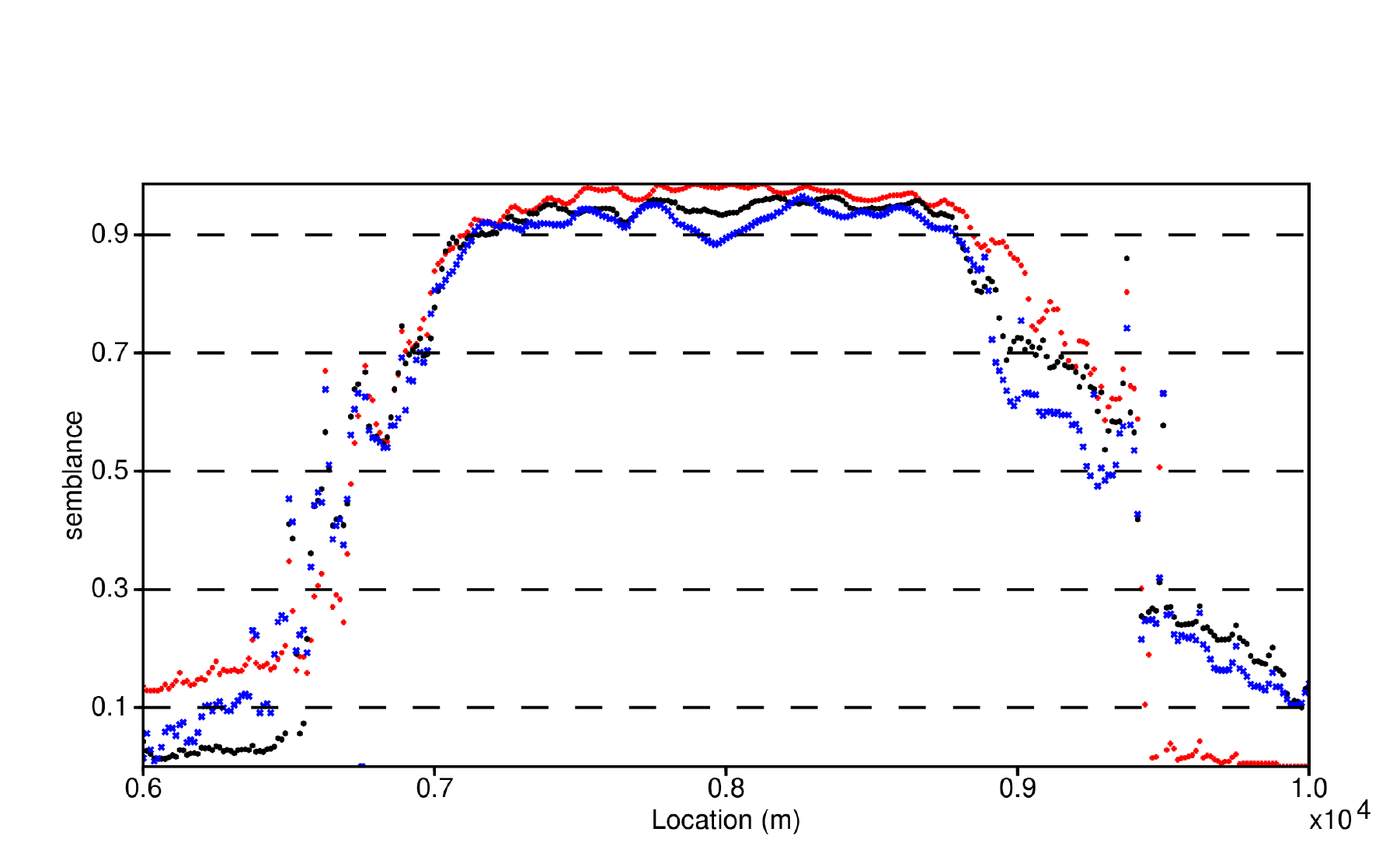}
\caption{semblances (reservoir base, monitor survey) true(red), initial(blue) and computed(black) velocities}
\label{fig:fig10}
\end{figure}

\begin{figure}
\captionsetup{justification=justified, singlelinecheck=off}
\centering
\includegraphics[width=0.45\textwidth]{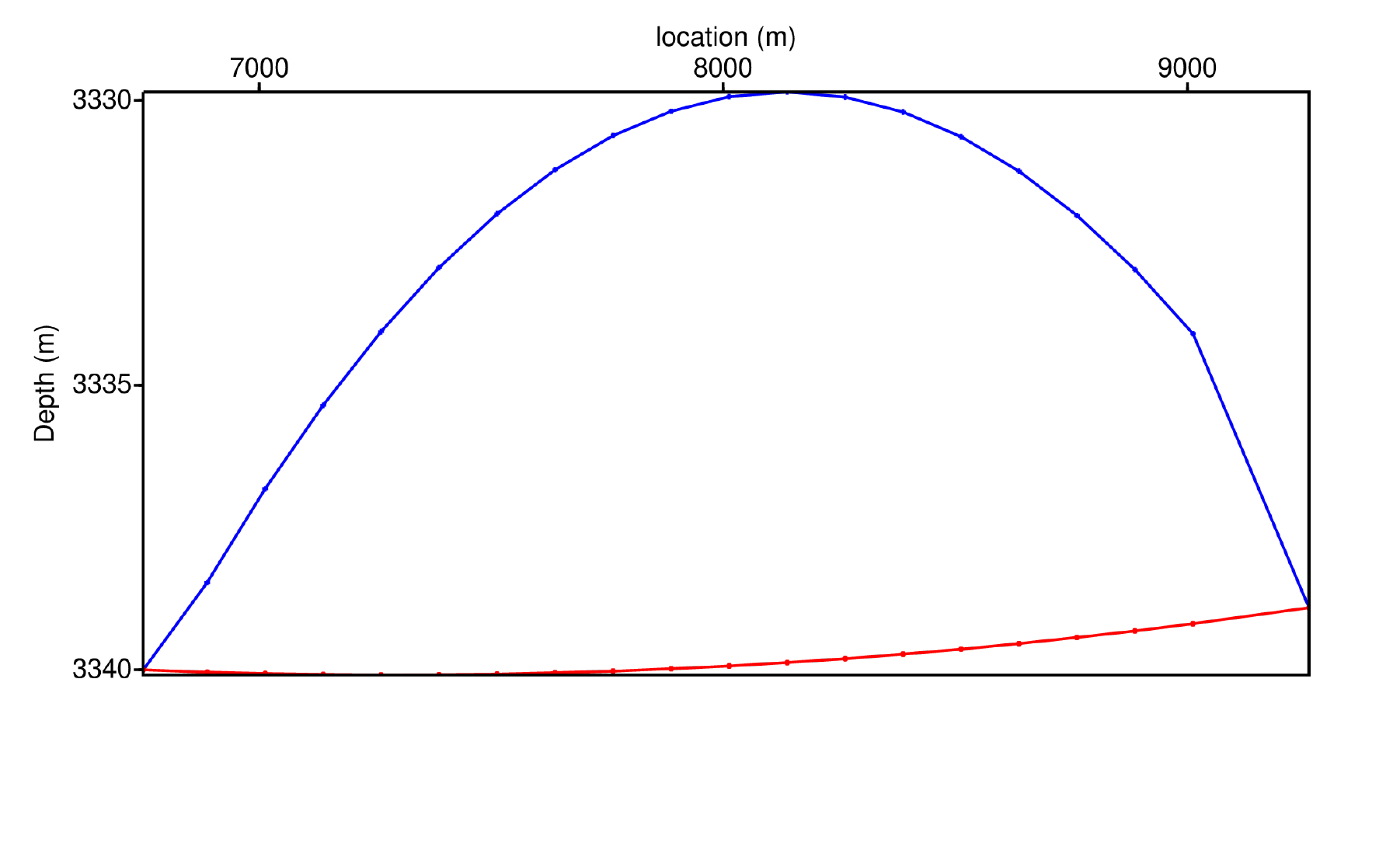}
\caption{compaction for the reservoir base: initial position (red) and after compaction (blue)}
\label{fig:fig11}
\end{figure}

Figure~\ref{fig:fig7} shows an approximation to the right fault of the Gulf of Mexico reservoir model described in (Oristaglio, 2016) with indicated reservoir boundaries, with the main difference that the reservoir does not consist of a sequence of turbidite fans separated by a shale layer. Here the reservoir thickness of the upper part is approximately 250m at a depth of 3340m, the layer is isotropic, again of constant velocity and for the calculated model response only primary reflected and diffracted events were considered. Again, two seismic surveys were assumed to be performed with 350 shots, 200 receiver/shot, maximum offset: 5000m, again the initial migration of the monitor survey was performed with the velocity model for the base survey. Figure~\ref{fig:fig8} shows the intermediate and far offset traces of this migration at location 7800m. The semblances for the velocity distributions in Figure~\ref{fig:fig10} are close to unity. They feature the true, initial and the computed velocities (in black) which are a better approximation to the true model (in red) than the initial one (Figure~\ref{fig:fig9}, in blue) (for the semblance calculations in Figure~\ref{fig:fig10} the computed velocities were augmented for this presentation with the values of the true velocity distribution outside the computational range $5000m < x < 10000m$; it can be observed that the decline in accuracy for the computed velocities for $x > 9000m$ correllates with the rapidly decreasing semblances in Figure~\ref{fig:fig10}). These results also  demonstrate the similarity of the semblances for different velocity distributions at this depth.\\
Again a simple example of reservoir compaction is demonstrated: a ficticious parabolic shift of the reservoir base with zero displacement at both ends is shown in Figure~\ref{fig:fig11} which was again assumed to be unknown for the velocity estimation. The model response was recalculated; again the computed velocity distribution was not altered by the simulated compaction. 

From these applications it may be concluded that the suggested approach appears to be a promising candidate for the determination of the velocity distribution of the formation of interest. Hopefully it will be possible to deal with other problems encountered in timelapse surveys in separate analyses, e.g. for deviations in horizontal positioning or for different noise conditions.

\section{Conclusions}
The presented approach was designed to determine the lateral variation of the formation velocity of interest for a seismic timelapse survey. It is supposed to be used for reservoir exploration and perhaps for the initial phase of CO2 sequestrations. It should be applicable under relatively low S/N ratios; local applications are possible. The applicability under different signal to noise ratios can be assessed by using model calculations before the actual analysis of the migrated monitor survey is performed. Applications have been dicussed for representative reservoir models exhibiting lateral varying velocity structures. Satisfactory approximations with respect to the true velocity distributions have been obtained. For the cases considered it has been demonstrated that the influence of compaction changes on the velocity calculations is small.

\section{Acknowledgements}

The Seismic Unix package has been used for processing and presentation purposes. 

\section{References}

Bednar J., 2005, A brief history of seismic migration: Geophysics, 70, 3MJ-20MJ\\
Born, M., 1985, Optik: Springer\\
Douma, M. and de Hoop, H., 2006, Explicit expressions for prestack map time migration in isotropic and VTI media and the
applicability of map depth migration in heterogeneous anisotropic media: Geophysics, 71, 513-528.\\
Guillaume, P., Lambare, G., Leblanc, O. et alii ,2008, Kinematics invariants: an efficient and flexible approach for velocity model building: 78 th SEG Annual International Meeting, Expanded Abstracts.\\
Hagedoorn, J., 1954, A process of seismic reflection interpretation: Geophysical Prospecting, 2, 85-127.\\
M. Khoshnavaz, H. Siahkoohi and A. Kahoo, 2021, Seismic velocity analysis in the presence of amplitude variations using local semblance, Geophysical Prospecting, 69, 1208–1217\\
Lambare, G., Herrmann, P., Toure, J., Suaudeau, E. and D. Lecerf, 2008, Computation of kinematic attributes for prestack time migration: SEG, 78 th Annual annual meeting, Expanded Abstracts, 2402-2406.\\
Lindsey, J.P. 1989, The Fresnel zone and its interpretative significance: The Leading Edge, 8, 33-39\\
Oristaglio, M., 2016, SEAM update: The Leading Edge, 35, 912-916\\
M. J. D. Powell. “The BOBYQA algorithm for bound constrained optimization without derivatives,” Department of Applied
Mathematics and Theoretical Physics, Cambridge, NA2009/06 (2009).\\
Robein, E., 2010, Seismic Imaging: EAGE.\\
Røste, T., Stovas, A., and Landrø, M., 2006, Estimation of layer thickness and velocity changes using 4D prestack seismic data,Geophysics, 71, S219-S234\\
Schneider, J., 1989, Specular prestack migration, 51th EAGE conference\\
Schneider, J., 2014, Aspects of residual moveout after downward continuation, 76th EAGE conference\\
Schneider, J., 2020, Kinematic aspects of residual prestack migration: arXiv:2003.12888 [physics.geo-ph]\\
Schneider, J., 2021, Application of the NIP wave theorem to PSDM and an approximation for RPSM to zero offset: arXiv:2111.09371 [physics.geo-ph]\\
Taner, M. and Koehler, F., 1969, Velocity spectra – digital computer derivation and applications of velocity functions.
Geophysics, 34, 859–881\\
Yilmaz, O, 2001, Seismic Data Analysis: SEG

\end{document}